\def\puncspace{\ifmmode\,\else{\ifcat.\C{\if.\C\else\if,\C\else\if?\C\else%
\if:\C\else\if;\C\else\if-\C\else\if)\C\else\if/\C\else\if]\C\else\if'\C%
\else\space\fi\fi\fi\fi\fi\fi\fi\fi\fi\fi}%
\else\if\empty\C\else\if\space\C\else\space\fi\fi\fi}\fi}
\def\SP{\let\\=\empty\futurelet\C\puncspace}

\def\h1{$h^{-1}$\SP}

\def\etal{{\it et al.\/}\ }

\def\lsim{~\rlap{$<$}{\lower 1.0ex\hbox{$\sim$}}}
\def\gsim{~\rlap{$>$}{\lower 1.0ex\hbox{$\sim$}}}
\def\void#1{{}}
\documentclass{aa}

\usepackage{times}
\usepackage{mathptm}
\fontfamily{ptmcm}\selectfont

\usepackage{graphicx}
\begin{document}

   \thesaurus{06     
              (03.11.1;  
               16.06.1;  
               19.06.1;  
               19.37.1;  
               19.53.1;  
               19.63.1)} 
   \title{ESO Imaging Survey}

   \subtitle{VI. The last 12 square degrees}

\author {
C. Benoist\inst{1} 
\and L. da Costa\inst{1} \and L.F. Olsen\inst{1,2}
\and E. Deul\inst{1,3} \and T. Erben\inst{1,5}  \and M.D.
Guarnieri\inst{6} \and R. Hook\inst{7} \and M. Nonino\inst{1,8}
\and I. Prandoni\inst{9}
\and M. Scodeggio\inst{1} \and R. Slijkhuis\inst{1,3}
\and A. Wicenec\inst{1} \and S. Zaggia\inst{1,10} }

\institute{
European Southern Observatory, Karl-Schwarzschild-Str. 2, D--85748
Garching b. M\"unchen, Germany \and
Astronomisk Observatorium, Juliane Maries Vej 30, DK-2100 Copenhagen, 
Denmark \and
 Leiden Observatory, P.O. Box 9513, 2300 RA Leiden, The
Netherlands
\and Institut d'Astrophysique de Paris, 98bis Bd Arago, 75014 Paris,
France
\and Max-Planck Institut f\"ur Astrophysik, Postfach 1523 D-85748,
 Garching b.  M\"unchen, Germany 
\and Osservatorio Astronomico di Pino Torinese, Strada Osservatorio
20, I-10025 Torino, Italy
 \and
Space Telescope -- European Coordinating Facility, Karl-Schwarzschild-Str. 2, 
D--85748 Garching b. M\"unchen, Germany 
\and Osservatorio Astronomico di Trieste, Via G.B. Tiepolo 11, I-31144
Trieste, Italy  \and 
Istituto di Radioastronomia del
CNR, Via Gobetti 101, 40129 Bologna, Italy
\and Osservatorio Astronomico di Capodimonte, via Moiariello 15,
I-80131.  Napoli, Italy
}

   \offprints{}

   \date{Received ; accepted }

   \maketitle
    

   \begin{abstract} This paper presents the I-band data obtained by
the ESO Imaging Survey (EIS) over two patches of the sky, 6 square
degrees each, centered at $\alpha \sim 5^h40^m$, $\delta \sim
-24^\circ50^m$, and $\alpha \sim 9^h50^m$, $\delta \sim -21^\circ
00^m$. The data are being made public in the form of object catalogs
and, photometrically and astrometrically calibrated pixel maps. These
products together with other useful information can be found at
"http://www.eso.org/eis". The overall quality of the
data in the two fields is significantly better than the other two
patches released earlier and cover a much larger contiguous area. The
total number of objects in the catalogs extracted from these frames is
over 700,000 down to $I\sim23$, where the galaxy catalogs are 80\%
complete. The star counts are consistent with model predictions
computed at the position of the patches considered.  The galaxy counts
and the angular two-point correlation functions are also consistent
with those of the other patches showing that the EIS data set is
homogeneous and that the galaxy catalogs are uniform.

   \end{abstract}

%

\section{Introduction}

This paper presents data for the last two patches (C and D) of the sky
observed by the public ESO Imaging Survey (EIS), being carried out in
preparation for the first year of regular operation of VLT. The I-band
data reported here covers a total area of 12 square degrees, down to
$I\sim 23$, corresponding to two patches probing separated regions of
the sky, 6~square degrees each. The present work complements earlier
papers in the series (Nonino \etal 1998; paper~I, Prandoni \etal 1998;
paper~III) and completes the presentation of the data accumulated by
the EIS observations carried out in the period July 1997-March 1998 as
part of the wide-angle imaging survey originally described by Renzini
and da Costa (1997) and in paper~I.

The primary science goal for surveying patches C and D was to search
for and produce a list of distant galaxy cluster candidates that would
complement those of the other two patches (A and B) reported earlier
(Olsen \etal 1998a,b: paper~II and V), providing VLT targets nearly
year-round. Patches C and D were also selected to overlap with the
ongoing 92cm Westerbork Survey in the Southern Hemisphere (WISH) being
carried out in the region $-15^\circ < \delta <-30^\circ$ and $\vert b
\vert > 10^\circ$.  Originally, the EIS observations were expected to
be carried out in two passbands (V and I). However, because of time
constraints and the prospect of supplementing the EIS observations at
the NTT with the new wide-field imager for the 2.2m ESO/MPIA
telescope, preference was given to increase the area covered by the
I-band observations, more suitable for identifying distant clusters
with $z\gsim0.6$ (see paper~V). This decision allowed the full
coverage of the selected patches, yielding a total coverage of 12
square degrees.  Combined with the data for patches A and B the EIS
I-band data covers a total area of about 17 square degrees, currently
the largest available survey of its kind in the Southern Hemisphere.

The goal of the present paper is to describe the characteristics of
the I-band observations of patches C and D. In section 2, the
observations, calibration and the quality of the data are
described. In section 3, the object catalogs extracted from the images
are examined and compared with data from the other patches and other
data sets to comparable depth. Concluding remarks are presented in
section~4.

\section{Observations and Data Reduction} \label{obs}

The observations of patches C and D were carried out over several
months in the period November (December for patch D) 1997 to March
1998, using the red channel of the EMMI camera on the 3.5m New
Technology Telescope (NTT) at La Silla.  The red channel of EMMI is
equipped with a Tektronix 2046 $\times$ 2046 chip with a pixel size of
0.266 arcsec and a useful field-of-view of about $9' \times 8.5'$. The
observations were carried out as a series of overlapping 150 sec
exposures, with each position on the sky being sampled at least twice,
using the wide-band filter WB829\#797 described in paper~I, and for
which the color term relative to the Cousins system is small.

\begin{figure}[ht]
\resizebox{1.\hsize}{!}{\includegraphics{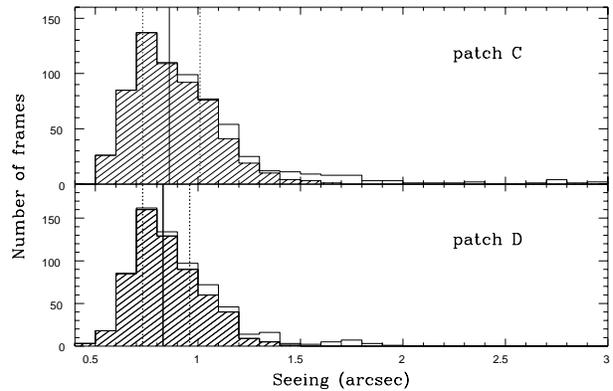}}        
\caption{The seeing distribution for the patches C and D obtained from all 
observed tiles (empty bars) and those actually accepted for the survey 
(shaded bars).}
\label{fig:seeing} \end{figure}

\begin{figure}[ht]
\resizebox{1.\hsize}{!}{\includegraphics{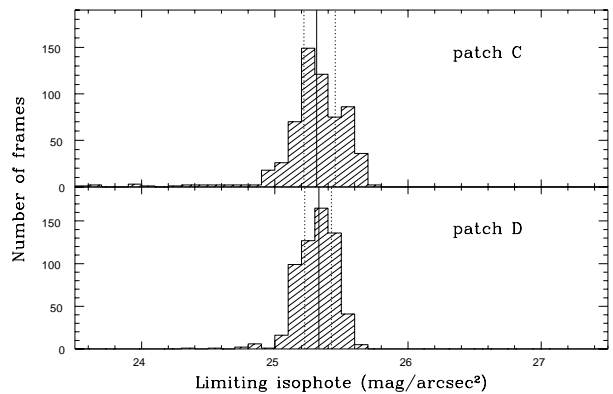}}               
\caption{The distribution of the limiting isophotes.}
\label{fig:limiso} \end{figure}

\begin{figure}[ht]
\resizebox{1.\hsize}{!}{\includegraphics{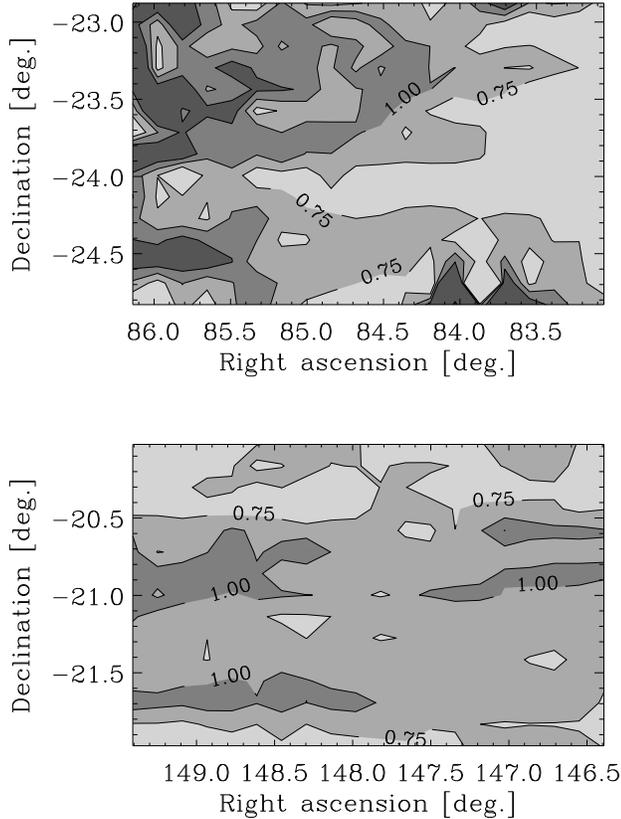}}
\caption{Two-dimensional distribution of the seeing as measured for
patches C (upper panel) and D (lower panel).}  \label{fig:seeing_cont}
\end{figure}

\begin{figure}[ht]
\resizebox{1.\hsize}{!}{\includegraphics{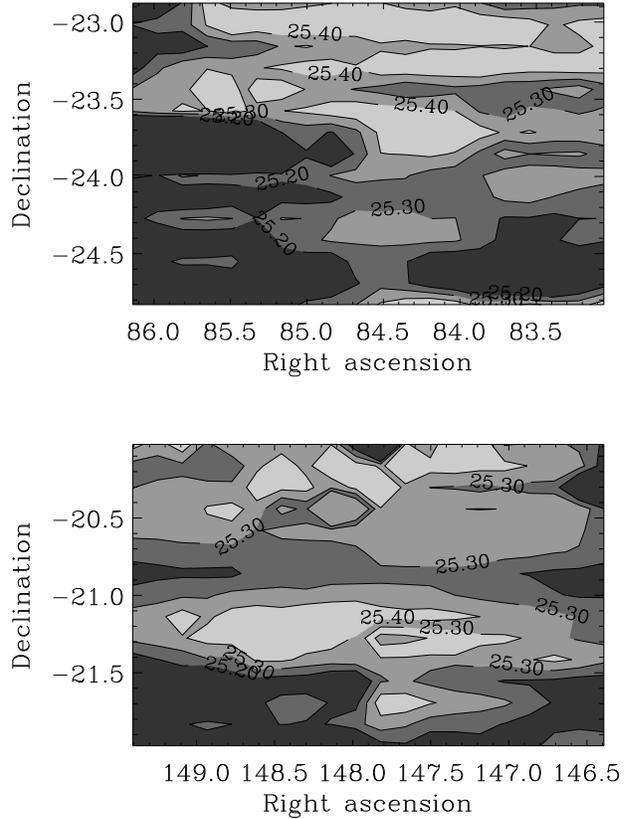}}
\caption{Two-dimensional distribution of the limiting isophote as
defined in the text estimated from the accepted even frames for
patches C and D.}  \label{fig:limiso_cont} \end{figure}

The data for patches C and D consist of 1348 frames but only 1203 were
accepted for final analysis, discarding 145 frames obtained in poor
seeing condition ($\gsim 1.5$ arcsec). The frames actually accepted
have a seeing in the range 0.5 to 1.6 arcsec, considerably better than
the data available for patches A and B obtained at the peak of El
Ni\~no.  Figure~\ref{fig:seeing} shows the seeing distribution of all
observed frames in each patch. For comparison the figure also shows
the seeing distribution of the accepted frames, with the vertical
lines in each panel indicating the median seeing and the quartiles of
the distribution. From the figure one finds that the median seeing for
both patches is sub-arcsec ($\sim 0.85$ arcsec) with only 25\% of the
area covered by frames with a seeing larger than 1~arcsec.  The good
quality of the observations can also be seen from
figure~\ref{fig:limiso} which shows the $1\sigma$ limiting isophote
within 1~arcsec for each patch.  Apart from one subrow in patch~C, in
both cases the limiting isophote is typically $\mu_I \sim$ 25.3 $\pm
0.1$ mag~arcsec$^{-2}$. The two-dimensional distributions of the seeing
and limiting isophote are shown in figures~\ref{fig:seeing_cont} and
\ref{fig:limiso_cont}. Comparison with similar distributions presented
in earlier papers (paper~I and III) shows that the data for patches C
and D are significantly better. Note that for each patch tables are
available listing the position of each accepted frame, its seeing,
limiting isophote and photometric zero-point and can be found at
``http://www.eso.org/eis''.

In late February 1998, a realignment of the secondary mirror was
carried out by the NTT team in an attempt to minimize the image
distortions seen in the upper part, especially the upper-right corner,
of the EMMI frames. Some frames for patch C and most of the frames in
patch D were observed with the new setup of the NTT. Examination of
the point spread function for these frames showed no significant
improvement in the quality of the images. This points out the need to
introduce a position-dependent estimator for the point-spread function
to assure uniformity in the star/galaxy separation across the
frame. This is particularly important for images observed under good
seeing conditions. In fact, examining the uniformity of the
classification as a function of position on the chip it is found that
there is a 10\% increase in the density of galaxies at the upper edge
of the chip, due to misclassifications, significantly larger than that
seen in paper~I.

In the last three runs (January-March) it was also noticed faint (at
the $1 \sigma$ level of the background noise) linear features aligned
along the east-west direction (perpendicular to the readout axis)
associated with moderately bright stars located in the lower half of the
CCD not previously seen. The cause for the these features are at the
present time unclear but are probably due to the electronic of the
old-generation CCD controller of EMMI, when used in a dual-port
readout mode. These affects two-thirds of the patch C frames and
essentially all the patch D frames. These light trails occur randomly
in the patch and there is no obvious way of correcting for them a
priori. An important consequence of this problem is that it leads to a
localized increase in the detection of low-surface brightness objects
over a range of magnitudes (typically $I\sim 20-21$) which can have a
significant impact in the cluster detection algorithm (Scodeggio \etal
1998, paper~VII).  This is unfortunate because both patches C and D
are located at lower galactic latitudes ($\vert b \vert \sim 25$) with
almost an order of magnitude larger density of stars than the previous
patches.

\begin{figure}[ht] 
\resizebox{0.8\hsize}{!}{\includegraphics{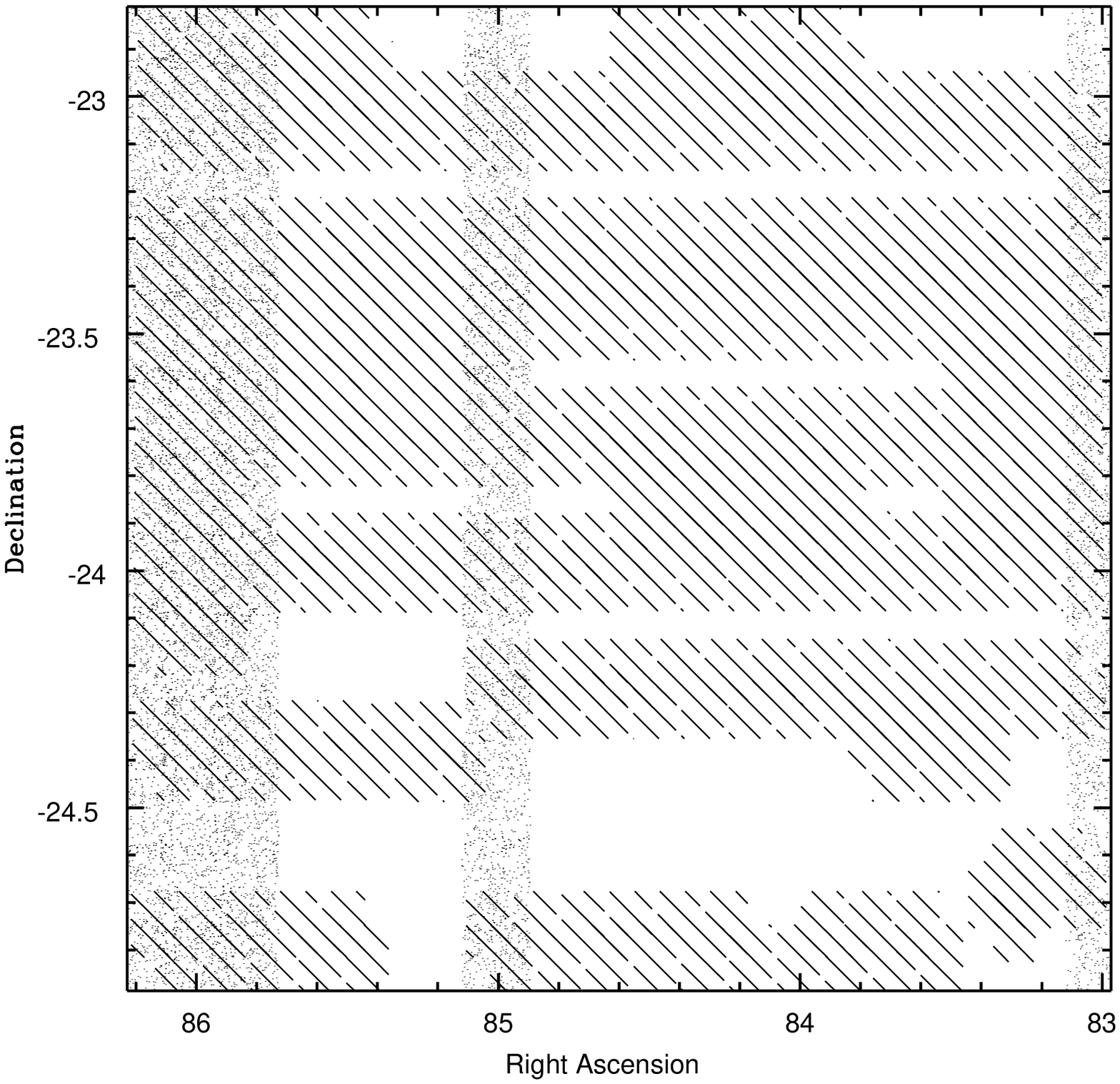}} 
\resizebox{0.8\hsize}{!}{\includegraphics{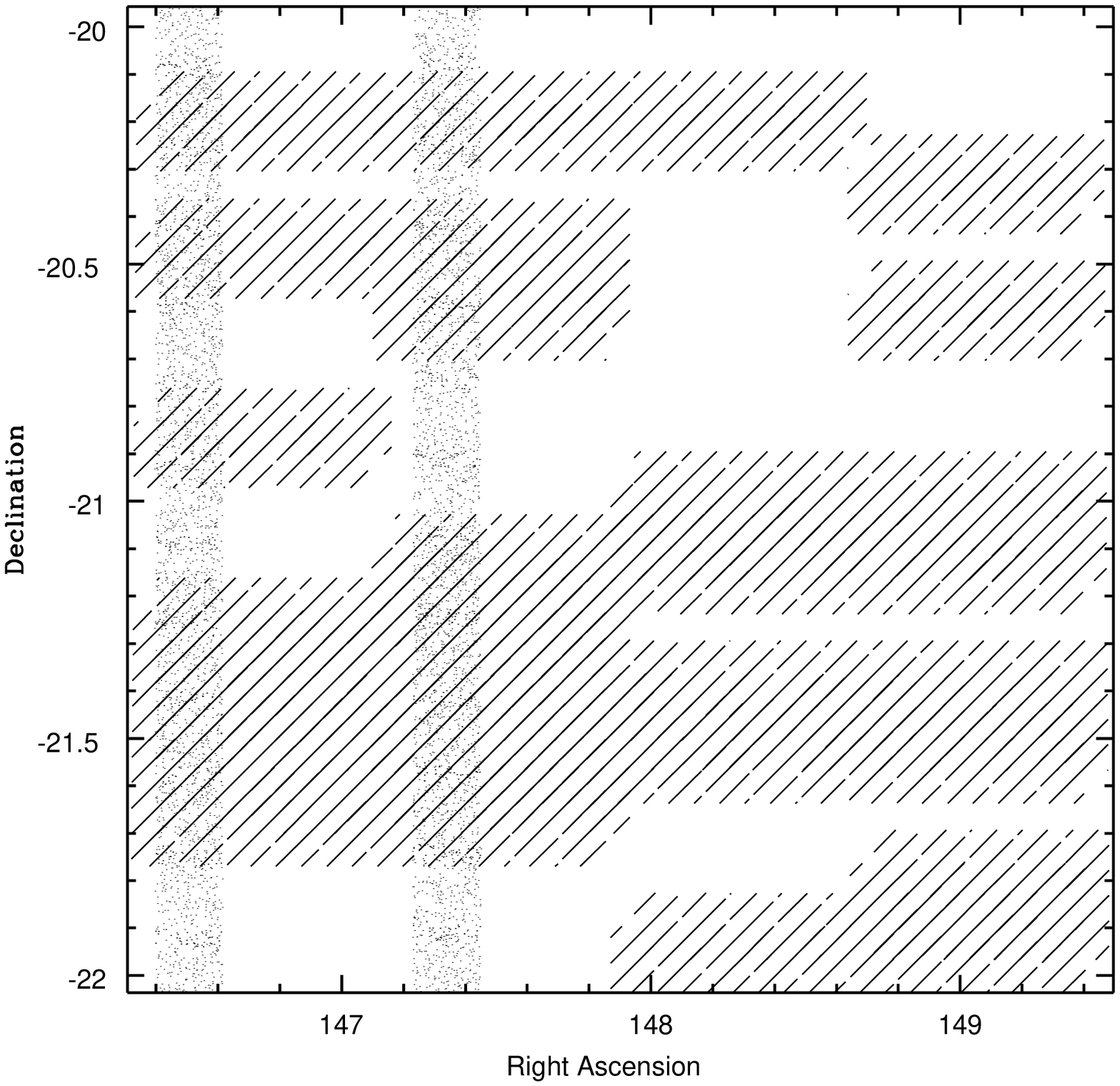}} 
              \caption{Overlap of DENIS strips that cross the the
surveyed area of patches C (top) and D (bottom).  The hatched area
represents regions containing EIS frames observed under photometric
conditions.}
\label{fig:overlaps} \end{figure}

The photometric calibration of the patch was carried out, as described
in papers~I and III, by determining a common zero-point for all frames
from the solution of a global least-squares fit to all the relative
zero-points, constraining their sum to be equal to zero. The absolute
zero-point was determined by a simple zero-point offset determined
from the common zero-point of all frames observed in photometric
conditions.  There are 340 and 290 such frames, covering about 80\%
and 60\% of the surveyed area, in patches C and D, respectively (see
figure~\ref{fig:overlaps}). The zero-points for these frames were
determined using a total of 10 fields containing of the order of 45
standard stars taken from Landolt (1992 a,b), observed in 10 nights
for patch C and in 11 nights for patch D. Altogether 215 independent
measurements of standards in the three passbands were used in the
calibration.  Comparison with external data suggests that a zero-point
offset provides an adequate photometric calibration for the entire
patch.

In order to check the photometric calibration and the uniformity of the
zero-points, strips from the DENIS survey (Epchtein
\etal 1996) crossing the surveyed area are  used.  The regions of
 overlap of these data are shown in Figure~\ref{fig:overlaps}, which
shows that there are five strips crossing patch C and two strips
crossing patch D. In the figure the regions observed under photometric
conditions are also indicated. Comparison of this figure with their
counterparts presented in papers I and III, clearly shows that the
data for patches C and D are of superior quality, with a much larger
fraction of frames taken under photometric conditions.

\begin{figure}[ht]
\resizebox{0.8\hsize}{!}{\includegraphics{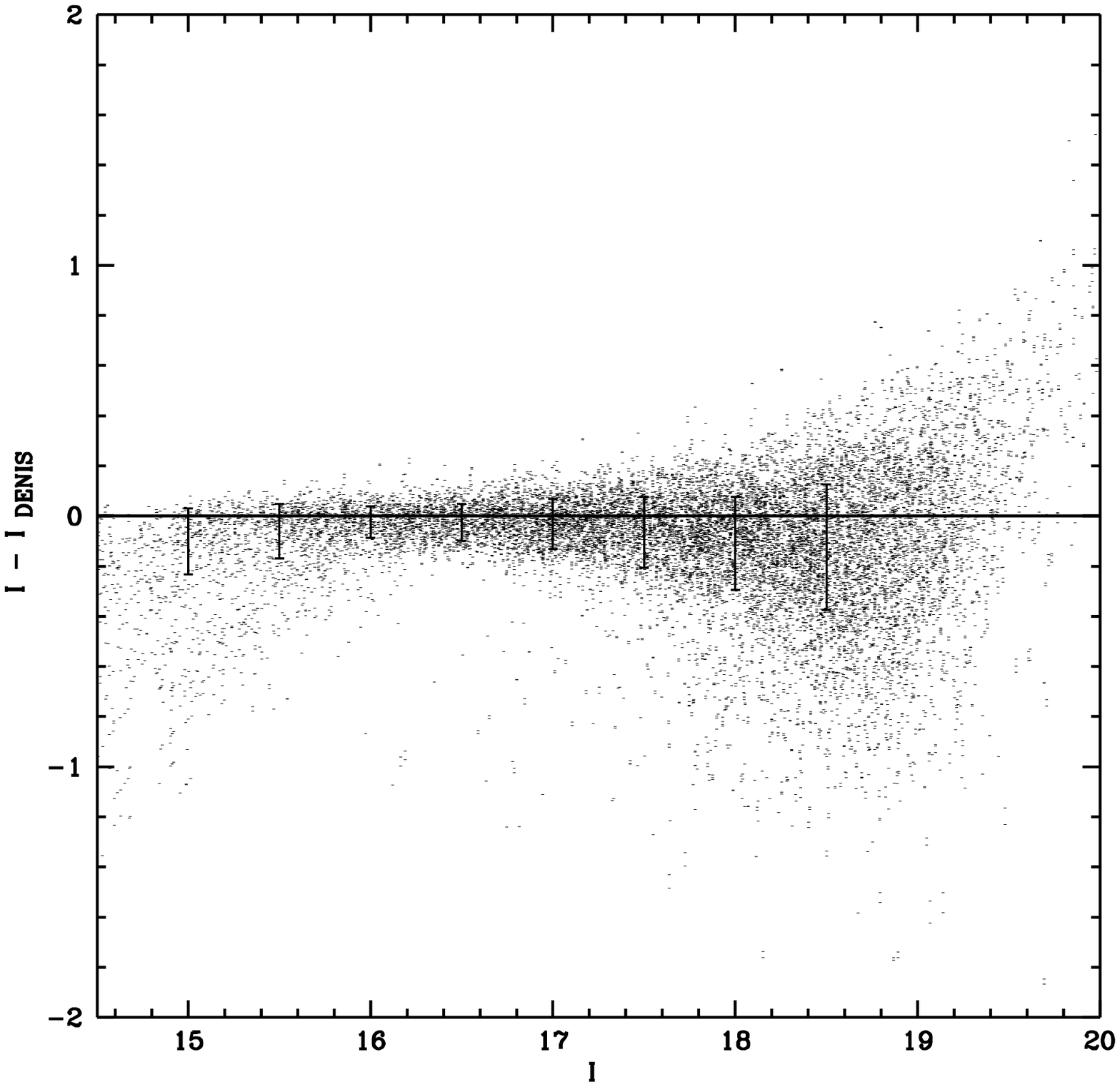}}
\resizebox{0.8\hsize}{!}{\includegraphics{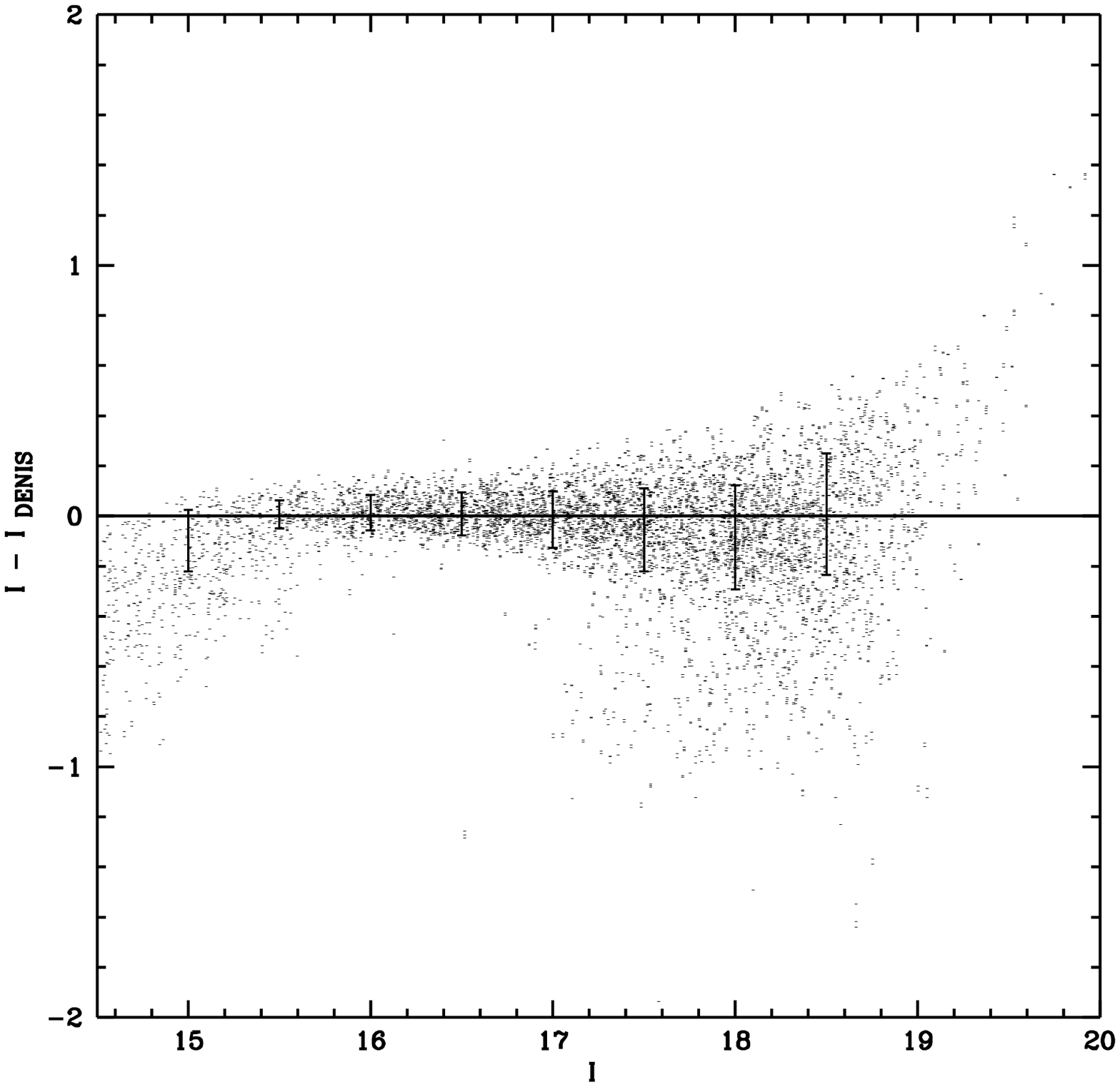}}               
\caption{Comparison of the EIS I-band magnitudes with those measured
by DENIS for the two strips that overlap patch C (top) and patch D
(bottom). Also shown are the mean and the rms in 0.5~mag bins. }
\label{fig:denis}
\end{figure}

In order to investigate possible systematic errors in the photometric
zero-point over the scale of the patch, the EIS catalogs were compared
with object catalogs extracted from the DENIS strips that cross the
survey regions (see figure~\ref{fig:overlaps}). Comparison of the
catalogs allows one to investigate the variation of the zero-point
over the patch. The results are shown in figure~\ref{fig:denis}.  The
domain in which the comparison can be made is relatively small because
of saturation of objects in EIS at the bright end ($I\sim16$) and the
shallow magnitude limit of DENIS ($I\sim18$). Still, within the two
magnitudes where comparison is possible one finds a roughly constant
zero-point offset of less than 0.02~mag for both strips and a scatter
of $\sim 0.2$ mag that can be attributed to the errors in the DENIS
magnitudes (Deul 1998).

\section{Data Evaluation}

\begin{figure*}[ht]
\resizebox{0.75\textwidth}{!}{\includegraphics{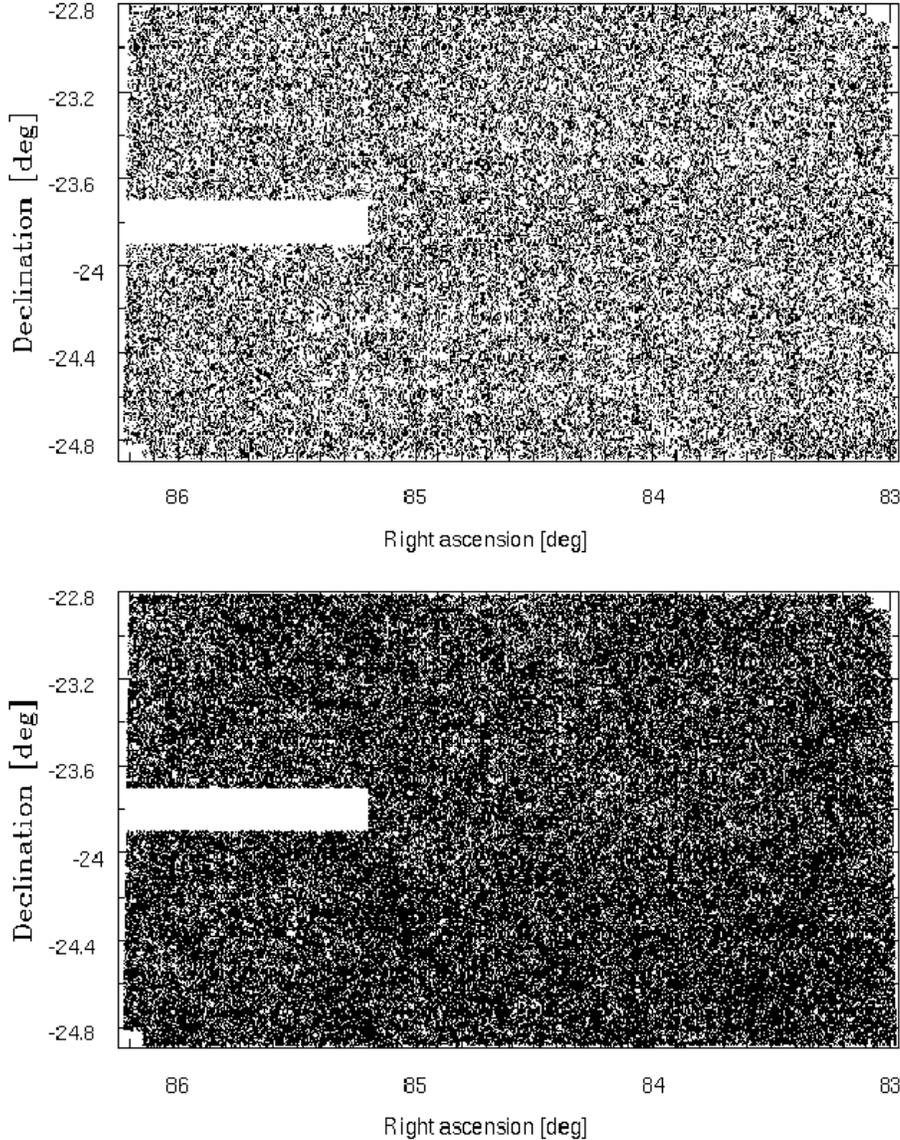}}
\caption{The projected distribution of stars (upper panel, $I\le
21.5$) and galaxies (lower panel, $I\le 22.5$) from patch C.}
\label{fig:pC_visu} \end{figure*}

\begin{figure*}[ht]
\resizebox{0.75\textwidth}{!}{\includegraphics{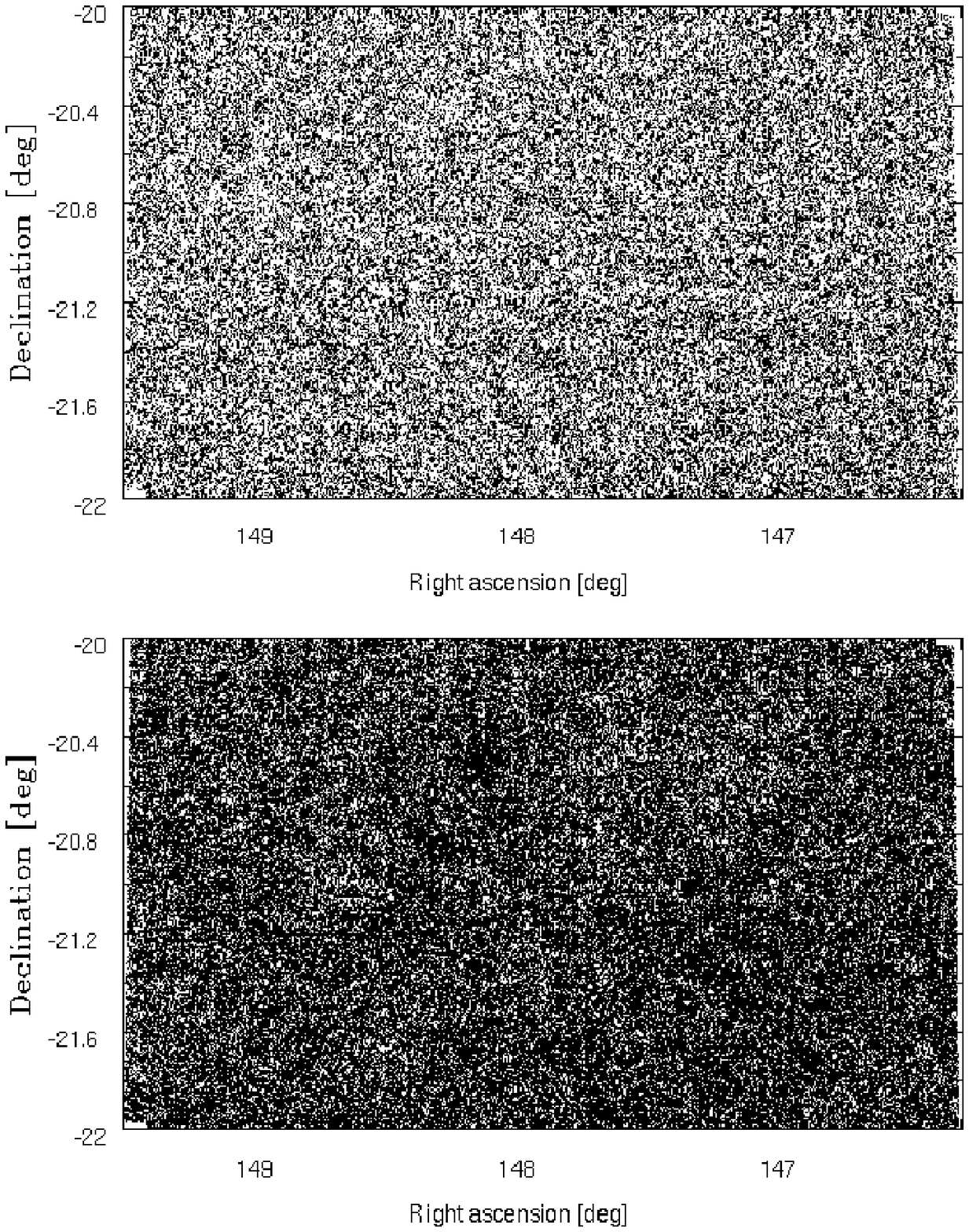}}               
\caption{Same as figure~\ref{fig:pC_visu} for patch D.}
\label{fig:pD_visu} \end{figure*}

In order to evaluate the quality of the data simple statistics
computed from the object catalogs extracted from the images are
compared in this section with model predictions and other data sets.
The catalogs derived from individual frames are used to generate the
even, odd and best seeing catalogs, described in earlier papers.  The
spatial distribution of stars and galaxies, defined using similar
star/galaxy classification criteria as in previous papers of the
series, are shown in figures~\ref{fig:pC_visu} and \ref{fig:pD_visu}
down to $I\sim 21.5$ and $I\sim22.5$ for stars and galaxies,
respectively.  The latter corresponds roughly to the completeness
limit of the object catalog. This limit was established using the
object catalog extracted from the co-addition of images of a reference
frame taken periodically during the observations of a patch. Note that
because of the much better seeing star/galaxy classification is
possible down to $I \sim 22$ and the completeness is about 0.5 mag
deeper. Some improvement in the classification is expected from a new
estimator being implemented in SExtractor based on a
position-dependent PSF fitting scheme currently being tested. 
This new version should also improve the uniformity of the
classification across the chip.

The distribution of the stars and galaxies shown in
figures~\ref{fig:pC_visu} and \ref{fig:pD_visu} is remarkably
homogeneous and considerably better than those seen in the previous
EIS patches due to the much better observing conditions.  This is true
except for a small region of about 0.2 square degrees in patch C which
has been removed, as indicated in figure~\ref{fig:pC_visu}. The only
problem seen with the galaxy catalogs in these patches is the presence
of several relatively thin linear features clearly seen at high
resolution (see EIS release page). These features are a consequence of
the electronic problem mentioned above and are not easily corrected
for at the image level.

In order to evaluate the data the general properties of the extracted
object catalogs are investigated and compared with model predictions
and other data sets.  Note that patches C and D are located at lower
galactic latitude and the number of stars is considerably larger. In
addition, the seeing is considerably better than in previous patches.
Therefore, it is of interest to re-evaluate the overall performance of
the EIS pipeline reduction under these new conditions.

\begin{figure}[ht]
\resizebox{1.\hsize}{!}{\includegraphics{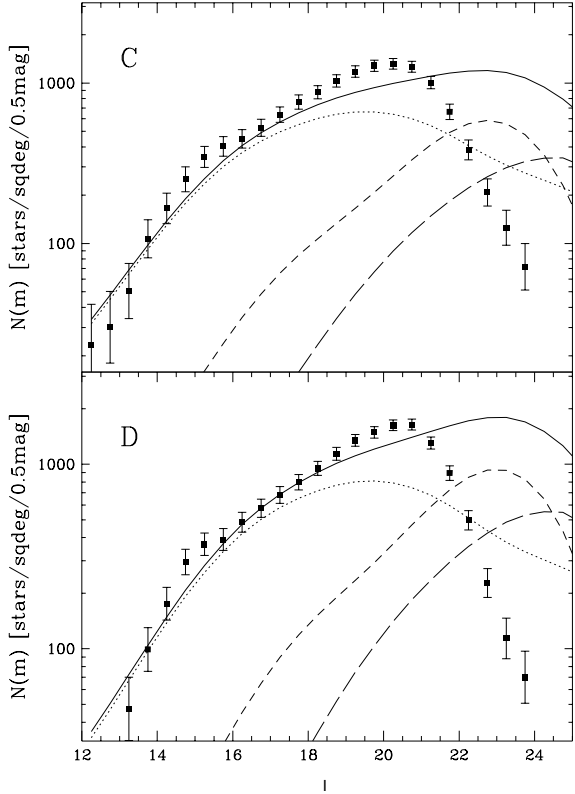}} \caption{The
differential star counts versus the I-magnitude for patches C and D
compared to model predictions (see
text). The dotted line represents the disc, the dashed line the thick
disc, the long-dashed line the halo, and the solid line the
sum of the three.}  \label{fig:star_counts} \end{figure}

Figure~\ref{fig:star_counts}, shows the comparison of the star counts
for patches C and D derived using the stellar sample extracted from
the object catalogs, with the predicted counts based on a galactic
model composed of an old-disk, a thick disk and a halo.  The
star-counts have been computed using the model described by M\'endez
and van Altena (1996), using the standard parameters described in
their Table~1 and an $E(B-V)$ of 0.015 and 0.010 for patches C and D,
respectively. It is important to emphasize that no attempt has been
made to fit any of the model parameters to the observed counts. The
model is used solely as a guide to evaluate the data. As can be seen
there is a good agreement at bright magnitudes ($ I \lsim 19$), but
the observed counts show an excess at fainter magnitudes ($18 <I<20$).
Even though it is unlikely that this excess is due to misclassified
galaxies at these relatively bright magnitudes, a better agreement can
be achieved if a higher stellarity index is assumed. On the other
hand, it is also possible that the model underestimates the
contribution of the thick-disk which makes a significant contribution
in this magnitude range.  The steep drop in the stellar counts beyond
$I\sim 21$ is partially due to the relatively high stellarity index
adopted, which was chosen to minimize the losses of galaxies. By
adopting a stellarity index of 0.5 the drop in the counts may be avoided
down to $I \lsim 21.5$. However, at these magnitudes and this value of
the stellarity index contamination by galaxies may be significant.
Another potential problem at these faint magnitudes is the
misclassification of stars as a consequence of the distortion effects
in EMMI, that can have some impact for images taken in good seeing
conditions.

\begin{figure}[ht]
\resizebox{1.\hsize}{!}{\includegraphics{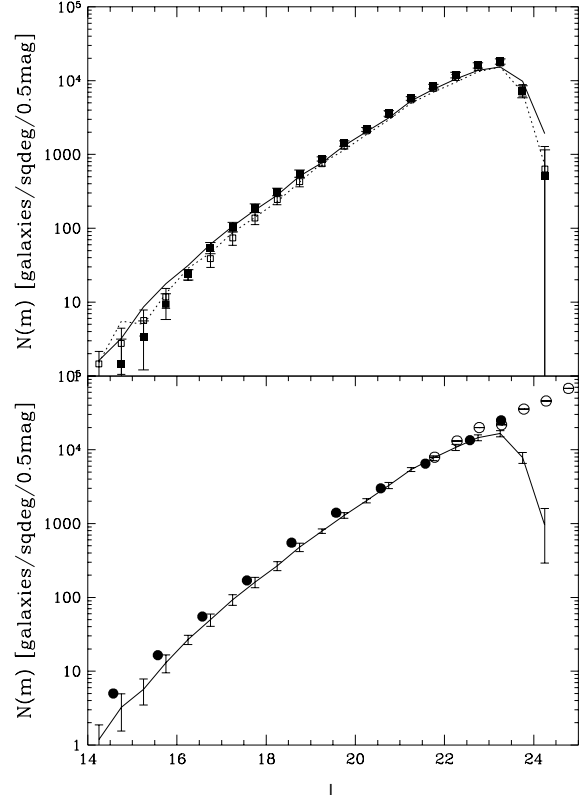}}               
\caption{Internal and external comparison of the EIS galaxy
counts. Upper panel shows the counts for patch C (open squares) and
patch D (filled squares), compared to the counts in patch A (solid
line) and patch B (dotted line). Lower-panel shows the average counts
for all EIS patches and the counts obtained by Postman \etal (1998)
(filled circles) and HDF (open circles).}
\label{fig:gal_counts} \end{figure}

In order to evaluate the depth of the galaxy samples, galaxy counts in
patches C and D are compared with those of previous patches in
figure~\ref{fig:gal_counts}. There is a remarkable agreement among the
counts derived for the different patches, indicating that the
identification of galaxies has not been affected by the observations
at lower galactic latitudes. The galaxy counts obtained from the
different patches have been combined to compute the mean galaxy counts
and the variance.  This is also shown in figure~\ref{fig:gal_counts}
where it is compared to other ground-based counts (Postman \etal 1998)
and those from HDF (Williams \etal 1996), appropriately converted to
the Cousins system (see paper~III).  As can be seen the EIS galaxy
counts agree extremely well with the ground-based data covering
comparable area over the entire magnitude range down to $I \sim 23$
and with the bright end of the HDF counts. The excellent internal and
external agreement of the I-band galaxy counts serves as a
confirmation of the reliability of the EIS galaxy catalogs.
Extraction from co-added images should allow reaching about 0.5~mag
deeper.

One way of examining the overall uniformity of the galaxy catalogs is
to use the two-point angular correlation function, $w(\theta)$, as
departures from uniformity should affect the correlation function
especially at faint magnitudes. The latter should be sensitive to
artificial patterns, especially to the imprint of the individual
frames, or possible gradients in the density over the field, which
could result from large-scale gradients of the photometric zero-point.
Note that any residual effect due to the improper association of
objects in the border of overlapping frames would lead to a grid
pattern (see the weight map in the EIS release page) that could impact
the angular correlation function.

\begin{figure}[ht]
\resizebox{1.\hsize}{!}{\includegraphics{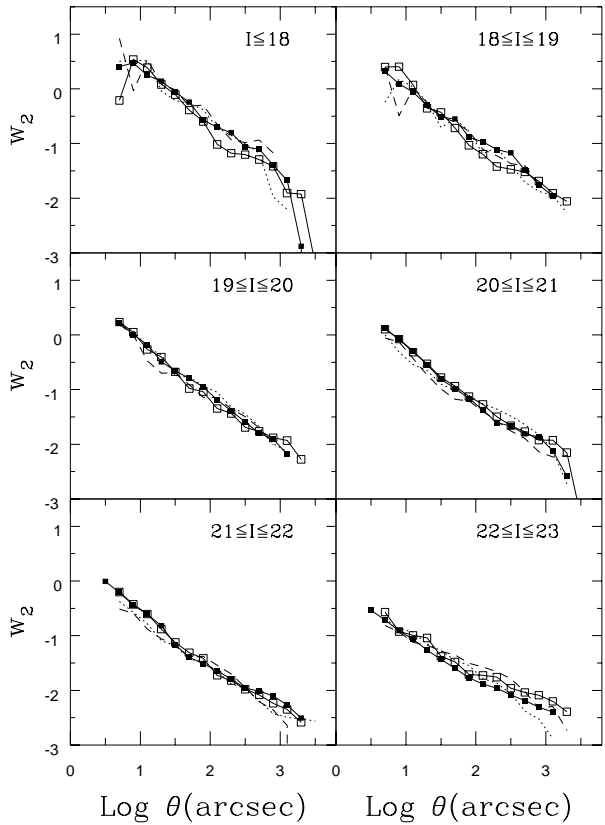}}               
\caption{Angular two-point correlation functions computed for patches C 
(open squares) and D (full squares). For comparison those obtained for
patches A (dotted line) and B (dashed line) are also shown.}
\label{fig:w} \end{figure}

Figure~\ref{fig:w} shows $w(\theta)$ obtained for different magnitude
intervals for both patches, using the estimator proposed by Landy \&
Szalay (1993). The calculation has been done over the entire area of
patch D and most of the area of patch C, with only one subrow (10
consecutive frames) removed according to the discussion above (see
section~\ref{obs}). For comparison, $w(\theta)$ computed for the other
patches are also shown (papers I and III) from which the cosmic
variance can be evaluated directly from the data. As can be seen there
is a remarkable agreement for all the magnitude intervals
considered. Moreover, the larger contiguous area of patches C and D
allows to estimate the angular correlation function out to $\sim 1$
degree. In all cases $w(\theta)$ is well described by a power law
$\theta^{-{\gamma}}$ with $\gamma$ in the range 0.7-0.8. Note that for
patch B the results refer to the galaxy sample obtained after removing
the foreground cluster (see paper III).  In particular, there is no
evidence for any underlying pattern associated with the overlap of
different frames.  The effect on $w(\theta)$ was evaluated by carrying
out simulations by adding to the observed galaxy distribution a grid
pattern with different density contrast. It was found that for high
contrast this would lead to local depressions in the angular
correlation function on scales of half the size of the diagonal of the
grid and its multiples, with the depth of depression depending on the
relative density. None such features are seen further indicating the
uniformity of the derived galaxy catalogs. Finally, note that the good
agreement of $w(\theta)$ for the different patches confirms that the
observed small-scale linear features associated with the faint light
trails, mentioned in section~\ref{obs}, have very little impact in the
angular correlation function.

As shown in paper~III the dependence of the amplitude of the
correlation function on the limiting magnitude of the sample is
consistent with earlier estimates based on significantly smaller areas
and the recent results reported by Postman \etal (1998). These results
show that the EIS galaxy catalogs are spatially uniform and form a
homogeneous data set independent of the patch, yielding reproducible
results.

\void{
As a final note, it is important to emphasize that there is a
suggestion for a gradual change in the shape of $w(\theta)$ for faint
galaxies ($I \gsim 20$).}

Finally, note that even though a single power-law with a slope between
0.7-0.8 gives a reasonable fit for the correlation computed in all
magnitude bins, there is some indication that for fainter samples ($I
\gsim 21$) the angular correlation function may be better
represented by two distinct power-laws. On small scales ($\lsim 30"$)
the slope remains the same while on larger scales it becomes gradually
flatter. A similar behavior is seen in the $w(\theta)$ computed for
all four patches. This flattening seems to be consistent with earlier
claims by Campos \etal (1995) and Neuschaefer and Windhorst (1995)
using significantly smaller samples, and more recently by Postman
\etal (1998) with a sample of similar size to EIS but covering a
single contiguous area.

\section{Summary}

One year after the first observations, the full data set accumulated
by EIS is being made public in the form of astrometrically and
photometrically calibrated pixel maps and object catalogs extracted
from individual images. In addition, separate papers have presented
derived catalogs listing candidate targets for follow-up work. The EIS
data set consists of about 6000 science and calibration frames,
totaling 96 Gb of raw data and over 200 Gb of reduced images and
derived products. All the information regarding these frames are
maintained in a continuously growing database. Together with the
Science Archive Group a comprehensive interface has been built to
provide users with a broad range of products and information regarding
the survey.

From the verification of the object catalogs and their comparison
against model predictions and other observations, it has been found
that the extracted catalogs are reliable and uniform. When all patches
are included, the combined EIS galaxy catalog contains about one
million galaxies and it is by far the largest data set of faint
galaxies currently available in the Southern Hemisphere. The star
counts show a good agreement with current galactic models, especially
at high-galactic latitudes, and the galaxy counts agree remarkably
well with other ground-based observations as well as with the counts
derived from HDF. The data from the different patches seem to be
rather homogeneous, as strongly suggested from measurements of the
angular two-point correlation function which should be sensitive to
large-scale gradients in a patch or to relative offsets of the
photometric zero-points for the different patches.

As expected EIS-wide has provided large samples (50 to over 200
candidates) of distant clusters of galaxies (Olsen \etal 1998a,b,
Scodeggio \etal 1998) and of potentially interesting point sources
(Zaggia \etal 1998), more than adequate for the first year of
observations with VLT, the main goal of EIS. Some of the targets can
also be observed nearly year round.  In order to expedite the delivery
of the products all the results refer to single exposure frames as
discussed in the previous papers of the series. Even though
co-addition has been done for all the patches some problems have been
uncovered during the verification of the object catalogs extracted
from them and require further work. However, the samples already
public are sufficiently deep and large for programs to be conducted in
the first year of operation of the VLT. The results obtained from the
co-added images will become available before the VLT proposal
deadline.

This paper completes the first phase of EIS which will now focus on
the deep observations of the HDF-south ($\alpha=22^h,
\delta=-66^\circ$) and AXAF deep ($\alpha=3^h, \delta=-25^\circ$) fields. The
results presented so far show the value of a public survey providing
the community at large with the basic data and tools required to
prepare follow-up observations at 8-m class telescopes. The experience
acquired by EIS in pipeline processing, data archiving and mining will
now be transferred to the Pilot Survey, a deep wide-angle imaging
survey to be conducted with the wide-field camera mounted on the
ESO/MPIA 2.2m telescope.

\begin{acknowledgements}
We thank all the people directly or indirectly involved in the ESO
Imaging Survey effort. In particular, all the members of the EIS
Working Group for the innumerable suggestions and constructive
criticisms, the ESO Archive Group and the ST-ECF for their support.
We also thank the Denis consortium for making available some of their
survey data. The DENIS project development was made possible thanks to
the contributions of a number of researchers, engineers and
technicians in various institutes. The DENIS project is supported by
the SCIENCE and Human Capital and Mobility plans of the European
Commission under the grants CT920791 and CT940627, by the French
Institut National des Sciences de l'Univers, the Education Ministry
and the Centre National de la Recherche Scientifique, in Germany by
the State of Baden-Wurttemberg, in Spain by the DGICYT, in Italy by
the Consiglio Nazionale delle Richerche, by the Austrian Fonds zur
F\"orderung der wissenschaftlichen Forschung und Bundesministerium
f\"ur Wissenschaft und Forschung, in Brazil by the Fundation for the
development of Scientific Research of the State of S\~ao Paulo
(FAPESP), and by the Hungarian OTKA grants F-4239 and F-013990 and the
ESO C \& EE grant A-04-046.  Our special thanks to the efforts of
A. Renzini, VLT Programme Scientist, for his scientific input, support
and dedication in making this project a success. Finally, we would
like to thank ESO's Director General Riccardo Giacconi for making this
effort possible in the short time available.

\end{acknowledgements}


\begin{thebibliography}{}


\bibitem{} Campos, A. et al. 1995, Clustering in the Universe, Maurogordato et
al. Eds., pp. 403-406. 

\bibitem{} Deul \etal 1998, {\it in preparation}

\bibitem{} Deul, E., 1998, {\it private communication}

\bibitem[Epchtein \etal 1996]{denis} Epchtein, N., \etal, 1996, The Messenger 87, 27

\bibitem{} Landolt, A.U., 1992a, AJ, 104, 340

\bibitem{} Landolt, A.U., 1992b, AJ, 104, 372

\bibitem{} Landy, S.D. \& Szalay, A. 1993, ApJ, 494, 1


\bibitem{} M\'endez R.A., and van Altena W.F. 1996, AJ, 112, 655

\bibitem{} Neuschaefer, L.W., \& Windhorst, R.A., 1995, ApJ, 439, 14

\bibitem{} Nonino, M. \etal, 1998, A\&A, {\it in press} (paper~I)

\bibitem{} Olsen, L.F. \etal, 1998a, submitted A\&A; astro-ph/9803338 (paper~II)

\bibitem{} Olsen, L.F. \etal, 1998b, submitted A\&A; astro-ph/9807156 (paper~V)

\bibitem{} Postman, M. \etal 1997, A\&AS, 191, 1903P

\bibitem{} Prandoni, I. \etal, 1998, submitted to A\&A;
astro-ph/9807153 (paper~III) 

\bibitem{} Renzini, A., \& da Costa, L.N. 1997, The ESO Messenger, 
No 87, p. 23  

\bibitem{} Scodeggio, M. \etal, 1998, submitted A\&A (paper~VII)

\bibitem{} Williams \etal 1996, AJ, 112, 1135

\bibitem{} Zaggia, S. \etal, 1998, submitted A\&A; astro-ph/9807152 (paper ~IV)

\end{thebibliography}
\end{document}